\documentstyle[12pt,aasms4]{article}   

\journalid{337}{15 January 1989}
\articleid{11}{14}
\slugcomment{}

\begin{document}
\title{The SiC problem: astronomical and
meteoritic evidence}
\author{A. K. Speck\altaffilmark{1}, A. M. Hofmeister\altaffilmark{2}, 
\& M. J. Barlow\altaffilmark{1}}

\altaffiltext{1}{Department of Physics \& Astronomy,
University College London, Gower Street, London WC1E 6BT, U.K.}
\altaffiltext{2}{Department of Earth \& Planetary Science, Washington 
University, St Louis, MO 63130,USA}

\begin{abstract}

Pre-solar grains of silicon carbide found in meteorites and interpreted as
having had an origin around carbon stars from their isotopic composition,
have all been found to be of the $\beta$-SiC polytype. Yet to date fits to
the 11.3-$\mu$m SiC emission band of carbon stars had been obtained only
for $\alpha$-SiC grains. We present thin film infrared (IR) absorption
spectra measured in a diamond anvil cell for both the $\alpha$- and
$\beta$- polymorphs of synthetic SiC and compare the results with
previously published spectra taken using the KBr matrix method. We find
that our thin film spectra have positions nearly identical to those
obtained previously from finely ground samples in KBr.  Hence, we show
that this discrepancy has arisen from inappropriate `KBr corrections'
having been made to laboratory spectra of SiC particles dispersed in KBr
matrices. We re-fit a sample of carbon star mid-IR spectra, using
laboratory data with no KBr correction applied, and show that $\beta$-SiC
grains fit the observations, while $\alpha$-SiC grains do not. The
discrepancy between meteoritic and astronomical identifications of the
SiC-type is therefore removed. This work shows that the diamond anvil cell
thin film method can be used to produce mineral spectra applicable to
cosmic environments without further manipulation. 

\end{abstract}

\keywords{methods: laboratory --- infrared: ISM: lines and bands ---
infrared: stars --- stars: carbon}

\section{Introduction}              

Most of the solid material in the solar system is believed to have
originated as small particles that condensed in outflows from stars.
However, most solar system solids (predominantly silicates) have been
reprocessed and/or homogenized so extensively that even the most primitive
meteorite silicate samples no longer contain evidence of their origins.
But some types of dust particles in the solar system have not been
reprocessed and can potentially be associated with their stellar origin.
One such dust type, silicon carbide (SiC), is believed to be a significant
constituent of the dust around carbon-rich AGB stars (Gilman 1969;
Treffers \& Cohen 1974). Silicon carbide grains can be divided into two
basic groups: $\alpha$-SiC if the structure is one of the many hexagonal
or rhombohedral polytypes and $\beta$-SiC if the structure is cubic (e.g.,
Bechstedt et al. 1997). Silicon carbide grains exhibit a strong
mid-infrared feature between 10 and 12~$\mu$m, with the peak of the
$\beta$-SiC feature occurring about 0.4~$\mu$m shortwards of that for
$\alpha$-SiC.  Until now, the observed peak wavelengths of the SiC feature
in astronomical spectra have been interpreted as indicating $\alpha$-SiC
to be the dominant type of SiC around carbon stars (e.g. Baron et al.
1985; P\'{e}gouri\'{e} 1988; Groenewegen 1995; Speck et al. 1997a,b). In
fact, Speck et al. 1997a,b found no evidence of $\beta$-SiC in these
circumstellar environments. Silicon carbide grains found in meteorites
have isotopic compositions which imply that most of these grains were
formed around carbon stars, with small amounts forming around novae and
supernovae (see Hoppe \& Ott 1997; Ott 1993 and references therein). All
studies to date of meteoritic SiC grains have found them to be of the
$\beta$-type (Bernatowicz 1997). $\beta$-SiC will transform into
$\alpha$-SiC above 2100$^{\rm o}$C but the reverse process is
thermodynamically unlikely. There is therefore an apparent discrepancy
between the meteoritic and astronomical SiC-types, which has been
discussed in detail by Speck et al. (1997a,b). 


We present new infrared (IR) absorption measurements of thin films of
$\alpha$- and $\beta$-SiC created by compression in a diamond anvil cell. 
Unlike some other methods, a dispersive medium (such as potassium bromide; 
KBr) is not used. This relatively new approach is quantitative, if
sufficient care is taken to produce an appropriately thin and uniform
film, as shown by comparison of thin film spectra of various minerals to
reflectivity data from the same samples (Hofmeister 1995; 1997 and
references therein). Moreover, thin film spectra of garnets are nearly
identical to single-crystal absorption data acquired in a vacuum
(Hofmeister, 1995):  hence, thin film spectra can be applied to
astronomical data without further manipulation.  Our measurements strongly
suggest, through comparison of the new thin film data with previous IR
spectra collected for fine-grained KBr dispersions (in which the dust
particles are dispersed in a KBr pellet), that the ``matrix correction''
wavelength shift, invoked by Dorschner et al. (1978) and adopted by other
authors (e.g. Friedemann et al. 1981; Borghesi et al. 1985), should not be
applied to laboratory spectra of sub-micron grain size dispersions of SiC:
it was the use of this ``KBr correction'' which caused the above-mentioned
discrepancy between the SiC-types found in meteorites and around carbon
stars.  A companion paper (Hofmeister \& Speck, in preparation)  clarifies
of the roles of scattering, absorption, reflection, and baseline
correction in laboratory measurements, sheds light on problems associated
with the powder dispersion technique, and discusses the conditions
appropriate for the application of such data.

\section{Laboratory techniques and results for thin-film samples}

Single-crystals of $\alpha$-SiC were purchased from Alpha/Aesar 
(catalog no. 36224).  This specimen is 99.8\% SiC, consisting of hexagonal 
platelets of 50 to 250~$\mu$m in diameter and 5 to 15~$\mu$m thick.  Less than 
1\% of the platelets had an amber color, the remainder were pale grey.  All 
were transparent in the visible with smooth, highly reflective surfaces.  
Polycrystals of $\beta$-SiC were donated by Superior Graphite.  The purity of 
this sample is also 99.8\%.  One batch consisted of 1~$\mu$m powder, the other 
was a conglomerate of equant crystallites of up to 25~$\mu$m in size. For 
this study, only the gray crystals of $\alpha$-SiC
were examined.  

Mid-IR spectra were
obtained from 450 to 4000 cm$^{-1}$ (2.5-22.2~$\mu$m) at 2 cm$^{-1}$
($\sim$ 0.01~$\mu$m) resolution using a
liquid-nitrogen cooled HgCdTe detector, a KBr beam splitter and an evacuated
Bomem DA 3.02 Fourier transform interferometer. Thin films were created
through compression in a diamond anvil cell which was interfaced with
the spectrometer using a beam condenser. Type II diamonds were used.
Film thickness was estimated from the initial grain size, by the 
relative relief and color seen among the various films, and by the increase in 
grain diameter from the initial size during compression. Efforts were made to 
cover the entire
diamond tip (0.6 mm diameter) with an even layer of sample, but slight
irregularities in the thickness were inevitable.  Reference spectra were
collected from the empty DAC. Uncertainties in peak positions are related to
peak widths because the accuracy of the FTIR spectrometer is high, 
$\pm$0.01~cm$^{-1}$. For procedural details see Hofmeister (1997).

Spectra obtained from $\alpha$-SiC (Fig. 1a,b) have an intense, broad band
near 11.8~$\mu$m.  The peak position lies between the longitudinal optic mode
(LO) and transverse optic mode (TO) components
observed by Spitzer et al. (1959) and a shoulder is seen at the LO 
position. A shoulder also occurs at 12.2~$\mu$m. 
The sample thickness could not be precisely determined,
but was estimated to be sub-micron.
Spectra obtained from $\beta$-SiC (Fig. 1c-g) depend somewhat on
thickness.
For the thinnest films, of sub-micron thickness (Fig.~1c,d), a fairly
symmetric peak is found at 11.3 to 11.4~$\mu$m, and a weak shoulder exists
at 10.7~$\mu$m, consistent with excitation of the LO component.  Spectra
from thicker film samples, $\sim$ 1~$\mu$m in thickness from visual inspection
(Fig.~1e,f), have a peak at a similar position, with an asymmetric
increase in intensity on the short-wavelength side, and display additional
weak features. 
The 12.7~$\mu$m band is due to the TO feature. The weak, broad band at
13.4 microns is not an absorbance feature but is due to the Christiansen
effect which gives a minimum when the real part of the index of refraction
is unity (Hapke 1993). The asymmetry of the main peak is
due to the baseline rising towards the visible, probably a scattering effect
from the grain boundaries.  A spectrum from the thickest sample examined
($\sim$ 1~$\mu$m), has high absorbance values overall, with the Si-C stretching
peak superimposed (Fig.~1g).  The appearance of the peak is intermediate
between the peaks observed from the thin (Fig.~1d) and moderately thick
samples (Fig.~1e), in
that the main peak is symmetric but a weak subsidiary feature exists at 
12.7~$\mu$m.  Below 8~$\mu$m, the absorbance in Fig.~1g drops, rather
than increasing as in the other spectra, because of interference fringes in the
near-IR (not shown). These interference fringes indicate a distance of 5~$\mu$m,
inferred to be the separation of the
diamond anvils.   The peak positions of the $\beta$-SiC samples are
relatively independent of the thickness. No difference can be discerned
between the two samples of $\beta$-SiC (fine grain
size vs. mixed grain sizes).  Fig.~1c,g were made from a mixture of grain
sizes and Fig.~1d,e,f are from the 1~$\mu$m powder fraction.  Additional
spectra from both $\beta$- samples resembled those shown. The appearance
of the spectra are consistent with being due to pure absorption
for the thinnest samples (Fig.~1c,d) and absorption with minor reflection
for the thicker samples (Fig.~1e-g), given that the LO-TO coupling is
stronger in $\beta$-SiC than in $\alpha$-SiC (Hofmeister \& Speck,
in preparation).

\section{Comparison with dispersed-sample results}

KBr matrix spectra of $\beta$-SiC obtained by Borghesi et al. (1985) for
a fine grain size sample (mean diameter modeled by them as 0.02~$\mu$m,
average diameter observed in TEM as 0.12~$\mu$m) closely match our own
thin-film data, particularly the spectrum in Fig.~1e (shown in Fig.2a). The
greatest difference is that the TO mode appears as a shoulder, rather than a
separate peak. This difference is obviously due to sample thickness, because
the thinner film of Fig.~1d has a barely discernable shoulder at the TO
position.  The LO mode occurs as a weak shoulder in their dispersion
data.  Their uncorrected peak barycenter was at 11.4~$\mu$m,
the same as for our thin films.  The match of Borghesi et al.'s dispersion data
with Fig.~1e is
consistent with an estimated film thickness $<$0.1~$\mu$m. The $\beta$-SiC 
spectrum of  Papoular et al. (1998), with maximum absorbance of 0.4, has 
a peak at 11.5~$\mu$m, in agreement with previous results and Fig.~1.  Papoular 
et al. (1998) also present two unusual spectra of $\beta$-SiC consisting of
broad overlapping peaks at 10.9 and 12.2~$\mu$m. These positions are close to the TO and 
LO
components. The very high absorbance units of 1 and 2.5 for these samples
suggest over-loaded pellets.  For extreme concentrations of SiC (or large
thicknesses), light is reflected between the TO and LO modes: the scattering
in the pellet produces the dip in absorption.  Problems occur at high
absorption because the partial opacity induces a frequency dependent baseline.

For $\alpha$-SiC, the KBr-dispersion spectrum of Borghesi et al. (1985)'s 
smallest-grained (mean diameter modeled by them as 0.04~$\mu$m, average 
diameter observed in TEM as 0.16~$\mu$m) and purest sample (SiC-600) closely 
matches the spectrum of our thinnest film (Fig.~1a; comparison shown in 
Fig.~2b). Its peak 
position of 11.6~$\mu$m equals our result, given the experimental 
uncertainties. The positions of the  shoulders are comparable to the LO and TO 
positions (Spitzer et al. 1959). Their sample N 
is compromised by $\sim$ 10\% impurities (C and SiO$_{2}$).
Their SiC-1200 sample was 3-10 times larger grained, even for the ground and 
sedimented fraction, and is inappropriate for comparison.  The study by 
Friedemann et al. (1981) involved larger grain sizes, but yielded similar 
spectral profiles, with a slight shift of the peak position to
11.8~$\mu$m.

It is clear that the introduction of a KBr matrix wavelength
correction (e.g. Friedemann 1981) is incorrect, since the barycenter
peak for KBr dispersions with fine grain sizes and reasonably low
concentrations equals that of corresponding thin films, while the
peak shapes
are in excellent agreement.  For these ($<$0.1~$\mu$m) grain sizes or film 
thicknesses, bulk absorption rather than surface effects dominates in the 
vicinity of the intense peak. Only for extremely thick or large grain samples, $\sim$ 
1~$\mu$m, do the parameters of the dispersions 
differ from those of a bulk sample but the differences are due to internal 
scattering among the particulates and sample opacity leading to incorrect 
assumptions for zero transmission. This issue is discussed further by
Hofmeister \& Speck (in preparation). Similarly, the application of a KBr 
correction for silicates (Dorschner et al. 1978) is also problematic.  
Recent measurements by Colangeli et al. (1993, 1995) indicate minimal 
matrix effects for various silicates. Thin film data on the other hand
do not suffer from these problems.

\section{Implications for the SiC-type that best matches carbon star
spectra}

Having established that previous fits of laboratory spectra for SiC 
to astronomical spectra have been erroneous due to
the unnecessary application of a KBr correction factor, we have re-fitted
our own UKIRT CGS3 spectra of carbon stars (Speck et al. 1997a,b) without such
a correction. We used the same $\chi^{2}$--minimization routine  
described by Speck et al. (1997a,b) but
the Borghesi et al. (1985) data for $\alpha$-SiC (SiC-1200, SiC-600 and 
SiC-N) and for $\beta$-SiC, to which Speck et al. (1997a,b) applied 
the usual KBr correction, were used uncorrected this time.

A detailed discussion of the fitting procedure can be found in Speck et 
al. (1997a). The routine was used on the flux-calibrated spectra, over the 
whole wavelength range (7.5--13.5~$\mu$m). All attempted fits involved either a 
blackbody or a blackbody modified by a $\lambda^{-1}$ emissivity, together
with some form of silicon carbide. The results are listed in Table~1 and
representative sample fits are shown in Fig. 3. 
The $\chi^{2}_{R}$ values are the reduced $\chi^{2}$ values, given by
dividing the  $\chi^{2}$ value by the number of degrees of freedom. The
fitting routine was  unable to find fits for four of the spectra, those of
AFGL~341, AFGL~2699, V~Aql and  Y~CVn. However, these four spectra are unusual
in that they display a strong feature in the 
7.5-9.5~$\mu$m region (see Fig.~2 of Speck et al. 1997a), possibly
identifiable with $\alpha$:C--H hydrogenated amorphous carbon (Baron 
et al. 1987, Goebel et al. 1995), and 
need to be classified separately. Self-absorption by SiC grains is a
possibility in some cases (Speck et al. 1997a,b), so the fitting
procedure was repeated using 
either a blackbody or modified blackbody, together with
silicon carbide in both emission and absorption simultaneously. The results of this 
fitting are listed in Table~1: 13 of the 20 spectra that could 
previously be fitted by SiC in pure emission produced better fits
with self-absorption included. Four sources found to have SiC
absorption features by Speck et al. (1997a) were also re-fitted and the 
new results are shown at the bottom of Table~1. Two of these
four sources required interstellar silicate absorption as well as
circumstellar SiC absorption (see Speck et al. 1997a).

The results in Table~1 shows that there is an obvious predominance of the
$\beta$-SiC phase and that there is now no evidence for the $\alpha$-SiC
phase at all.  This is in contrast to previous attempts to fit the
astronomical SiC feature using similar, and in some cases the same, raw
laboratory data, but inappropriately corrected for the KBr dispersion.
Previous work found that the best fits were obtained with $\alpha$-SiC,
and had concluded that there was no unequivocal evidence for the presence
of any $\beta$-SiC. Without the KBr correction, $\beta$-SiC matches the
observed features, while $\alpha$-SiC does not. Thus there is now no
astronomical evidence for the presence of $\alpha$-SiC in the
circumstellar regions around carbon stars. While $\alpha$-SiC might exist
in small quantities, all observations to date are consistent with the
exclusive presence of $\beta$-SiC grains.  This resolves the past
discrepancy, reconciling astronomical observations and meteoritic samples
of silicon carbide grains.  Having confirmed that SiC grains observed
around carbon stars and those found in meteorites are of the same
polytype, further discrepancies need to be addressed. In particular, the
differences in grain sizes between astronomical models and meteoritic
grains merits attention (see Speck et al. 1997a,b for a detailed
discussion).

Furthermore, the current work has demonstrated that mineral spectra
produced using the DAC thin film method are directly applicable to
astrophysical contexts without further manipulation of the data.
It is now appropriate to use the DAC thin film method to produce more
mineral spectra of use to astronomers.

\vspace{0.3cm}
\noindent
Support for AKS was provided by the United Kingdom Particle Physics and
Astrophysics Research Council and by University College London. Support
for AMH was provided by Washington University.  We thank Chris
Bittner (Superior Graphite Co.) for providing samples, and Tom Bernatowicz
for 
suggesting this collaboration. This paper is dedicated to Dr. Chris Skinner, 
who died suddenly on October 21st 1997.

\vspace{10.0cm}

\figcaption{Fig.~1.--
Representative thin-film IR spectra.  (a-b) (lower): sub-micron
films of $\alpha$-SiC, which have an intense broad band at 11.70~$\mu$m
(a) and 11.87~$\mu$m (b). (c-g) (upper): $\beta$-SiC spectra acquired from
films increasing in thickness from $<$0.1~$\mu$m to $\sim$1~$\mu$m. For
clarity, plot (d) was offset by +0.2 absorbance units, and plot (b) by
+0.1 absorbance units. Absorbance clearly increases with thickness for
$\beta$-SiC. The positions of the longitudinal optic (LO) and transverse
optic (TO) modes of SiC are shown, as well as those of the overtone
and Christiansen feature (CF).}

\vspace{2.0cm}

\figcaption{Fig.~2.-- 
Our thin film spectra (solid lines) and the uncorrected Borghesi et al.
(1985) KBr dispersion data (dashed lines).}

\vspace{2.0cm}

\figcaption{Fig.~3.--
$\beta$-SiC (plus blackbody) fits to several UKIRT CGS3 carbon star
spectra. See text for details.}

\begin{table}
\caption{Results of the $\chi^{2}$-fitting for the
7.5--13.5~$\mu$m flux-calibrated carbon star spectra}
{\scriptsize
\begin{tabular}{l@{\hspace{0mm}}c@{\hspace{1mm}}c@{\hspace{1mm}}c
@{\hspace{1mm}}c@{\hspace{2mm}}c@{\hspace{2mm}}c}
Source  &  SiC type  &  $T_{BB}(K)$  &$T_{SiC}(K)$
&  $\tau_{SiC}$  &  $\tau_{9.7}$$^{\ast}$  &  $\chi^{2}_{R}$\\

IRAS 21489+5301$^{1}$&$\beta$-SiC   & 449  & 293 & ----- & --- &  0.515\\
IRC+10216$^{1}$   & $\beta$-SiC     & 511  & 230 & ----- & --- &  1.260\\
AFGL 5076$^{2}$   & $\beta$-SiC     & 557  & 298 & 0.137 & --- &  0.369\\
AFGL 2494$^{2}$   & $\beta$-SiC     & 516  & 383 & 0.167 & --- &  0.306\\
AFGL 3099$^{2}$   & $\beta$-SiC     & 726  & 329 & 0.242 & --- &  1.370\\
AFGL 5102$^{2}$   & $\beta$-SiC     & 650  & 355 & 0.161 & --- &  0.345\\
AFGL 2155$^{2}$   & $\beta$-SiC     & 734  & 288 & 0.235 & --- &  0.418\\
IRAS 02152+2822$^{2}$&$\beta$-SiC   & 548  & 519 & 0.223 & --- &  0.504\\
IRC+40540$^{2}$   & $\beta$-SiC     & 859  & 313 & 0.173 & --- &  0.508\\
AFGL 2368$^{1}$   & $\beta$-SiC     & 727  & 321 & ----- & --- &  1.772\\
V Hya$^{2}$       & $\beta$-SiC     &1129  & 393 & 0.211 & --- &  0.761\\
IRC+00365$^{2}$   & $\beta$-SiC     &1788  & 215 & 0.114 & --- &  2.316\\
CIT 6$^{2}$       & $\beta$-SiC     & 960  & 363 & 0.217 & --- &  1.294\\
IRC+50096$^{1}$   & $\beta$-SiC     & 940  & 455 & ----- & --- &  2.166\\
R For$^{1}$       & $\beta$-SiC     & 906  & 800 & ----- & --- &  1.237\\
R Lep$^{1}$       & $\beta$-SiC     &1284  & 573 & ----- & --- &  -----\\
UU Aur$^{2}$      & $\beta$-SiC     &2505  & 446 & 0.165 & --- &  1.105\\
V Cyg$^{2}$       & $\beta$-SiC     &2556  & 568 & 0.139 & --- &  1.014\\
CS 776$^{1}$      & $\beta$-SiC     & 993  & 576 & ----- & --- &  1.879\\
V414 Per$^{2}$    & $\beta$-SiC     &1102  & 920 & 0.177 & --- &  0.579\\
AFGL 3068$^{3}$         &  $\beta$-SiC    & 394  &  62 & 0.030 & ---- & 
0.092\\
IRAS 02408+5458$^{3}$   &  $\beta$-SiC    & 388  &  96 & 0.152 & ---- & 
1.686\\
AFGL 2477$^{\dag}$$^{3}$&  $\beta$-SiC    & 377  &  114& 0.073 & 0.104& 
0.419\\
AFGL 5625$^{\ddag}$$^{3}$& $\beta$-SiC    & 358  &  185& 0.097 & 0.113& 
0.306\\
\end{tabular}
\begin{tabular}{l}
1 Fits with pure emission only\\
2 Fits with self-absorbed net emission\\
3 Fits with self-absorbed net absorption\\
$^{\ast}$ $\tau_{9.7}$ is the optical depth at 9.7~$\mu$m\\
$^{\dag}$ also requires Trapezium interstellar silicate absorption\\
$^{\ddag}$ also requires $\mu$ Cep interstellar silicate absorption\\
\end{tabular}
}
\end{table}

\end{document}